\documentclass[twocolumn,showpacs,preprintnumbers,amsmath,amssymb]{revtex4}

\usepackage{graphicx}% Include figure files
\usepackage{dcolumn}% Align table columns on decimal point
\usepackage{bm}% bold math

\begin{document}

%\preprint{APS/123-QED}

\title{Electron-impact excitation and ionization of atomic calcium at intermediate energies}

\author{Oleg Zatsarinny}
\email[Electronic Address: ]{oleg_zoi@yahoo.com}
\author{Henry Parker}
\author{Klaus Bartschat}
\affiliation{Department of
Physics and Astronomy, Drake University, Des Moines, Iowa, 50311, USA}

\def\etal{{\it et al}.}

\date{\today}% It is always \today, today,

\begin{abstract}
We present a comprehensive study of electron collisions with calcium atoms by using the convergent \emph{B}-spline \emph{R}-matrix method. 
Elastic, excitation, and ionization cross sections
were obtained for all transitions between the lowest 39 physical states of calcium (except for $3p^6 4s5g\ ^{3,1}G$) 
up to the $3p^6 4s8s\ ^1S$ state,
for incident electron energies ranging from threshold to 100~eV.
A multi\-configuration Hartree-Fock method with non\-orthogonal term-dependent orbitals was employed to generate accurate 
wave functions for the target states. Close-coupling expansions including the 39 physical states plus 444 pseudo-target 
states of calcium were used to check the sensitivity of the results to coupling to the target continuum. 
The cross-section dataset obtained from the large-scale calculations is expected to be sufficiently accurate 
and comprehensive for most current modeling applications involving neutral calcium.
\end{abstract}

\maketitle

\section{Introduction}
\label{sec:Introduction}

Accurate atomic data for electron collisions with atoms and ions
are of importance in the modeling of a variety of astrophysical and laboratory plasmas.
Electron-atom collisions couple the electron gas to the radiation field
through atomic excitations and de-excitations, and thereby collision cross sections
are crucial in determining the state of matter from the observed radiation field.
This includes the investigation of the structure and
evolution of galaxies, gas clouds, stars, and other objects
that can be studied spectroscopically.
Particular applications include stellar atmospheres, where large deviations
from local thermal equilibrium (LTE) may be present.
Abundance ratios allow the study of possible stellar nucleo\-synthesis
mechanisms occurring in galaxies~\cite{Bark2016}.
Having a set of NLTE calculations for various elements
makes the comparison of the abundance ratios of the
atomic species in different types of stars more reliable.
Accuracy and completeness of the atomic data used is crucial in such modeling.

Calcium, one of the $\alpha$ elements, has observable lines along a broad range of wavelengths in
late-type stellar spectra. The first detailed study of calcium lines in solar spectra
was published by Holweger~\cite{Holw1972}.
The presence of Ca lines in both solar and stellar spectra was confirmed in many subsequent papers.
In particular, calcium lines were used for determinations of cool stellar atmospheric properties
and for the study of chromospheric activity in late-type stars~\cite{Thor2000,Chmi2000}.
Lines of both neutral and singly ionized Ca are observed even in most metal-poor stars~\cite{Chris2004}
and can be used for the determination of fundamental stellar parameters.

Electron scattering from calcium has been the subject of a large body of research, both experimental and theoretical.
Most of the studies, however, were devoted to elastic scattering and excitation of the resonance
transition to the $4s4p\ ^1P^o$ state. A comprehensive discussion of elastic \hbox{e-Ca} collisions at low energies
up to 4~eV is given in our previous paper~\cite{Ca2006}. In particular,
comparison with other theoretical predictions revealed that the numerical results can
be very model dependent. Comparison with various experimental data showed good agreement in some
cases, while discrepancies remained in others. It was not immediately clear, however,
whether these discrepancies were solely due to the omission of some important physics in the
computational model or whether experimental problems might be responsible as well.
The only experimental investigation of elastic \hbox{e-Ca} scattering at intermediate energies that we are aware of 
was carried out by Milisavljevic \emph{et al.}~\cite{Mil2005}. Employing a crossed electron-atom beam technique,
they measured the differential cross section (DCS) at incident energies of 10, 20, 40, 60 and 100~eV.
Comparison with numerous theoretical investigations, performed mainly with various modifications
of the optical model-potential (OMP) method, shows reasonably good agreement between the experimental
and calculated DCSs concerning their angular dependence, whereas the agreement between the absolute values 
is not as good. The absolute values obtained theoretically are generally larger than those measured.

The cross section for the allowed transition $4s^2 - 4s4p\ ^1P^o$ is probably the most accurately
known excitation cross section in Ca.
Apparent cross sections for electron-impact excitation of this resonance transition
were measured by Ehlers and Gallagher~\cite{EG1973}, whereas Milisavljevic
\emph{et al.}~\cite{Mil2004} obtained angle-differential cross sections
for this transition for incident energies between 10 and 100~eV.
This work prompted further theoretical efforts employing different methods and approaches.
Chauhan \emph{et al.}~\cite{Chauhan2005} used a first-order relativistic distorted-wave (RDW) model,
whereas Kawazoe \emph{et al.}~\cite{Kawazoe2005} applied 15-state, 21-state, and 24-state
non\-relativistic \emph{R}-matrix (close-coupling) models to the problem.
Later, Zatsarinny \emph{et al.}~\cite{Ca2007} used the \emph{B}-spline \emph{R}-matrix (BSR) method to 
study electron-impact excitation of the lowest four excited states of calcium, and afterwards
Fursa and Bray~\cite{Fursa2008} applied the convergent close-coupling (CCC) method to the 
study of the $^1P^o$ excitation of Ca.
These calculations with increasing level of sophistication improved the agreement with experiment.
In particular, the CCC calculations revealed that models that neglect the coupling to the target continuum
generally overestimate the cross section, even for this strong resonance transition.

The excitation of other states in Ca has been studied much less. Here we mention
the joint theoretical and experimental work of Shafranyoshy \emph{et al.}~\cite{Ca1997} devoted to
the study of electron-impact excitation of Ca atoms from the metastable state $4s4p\ ^3P^o$.
The limited \hbox{6-state} close-coupling calculations in this work did not achieve a comprehensive reproduction
of the experimental data, but they were able to explain the principal features. Among other works, we note
the RDW calculations by Muktavat \emph{et al.}~\cite{Muk2002} concerning excitation of the low-lying
$3d4s\ ^{1,3}D$ states in Ca. These states, along with the strong exchange transition to the $4s4p\ ^3P^o$ state,
were also considered in our previous calculations~\cite{Ca2007}, where considerable disagreement with the RDW
results was found. To our knowledge, the only attempt to provide a systematic dataset for electron-impact excitation of Ca
was under\-taken by Samson and Berrington~\cite{Sam2001} who presented excitation cross sections
and thermally averaged effective collision strengths for transitions from the $4s^2$ ground state to the next 10
states of Ca, based on a 22-state \emph{R}-matrix model.

Plasma modelling requires a comprehensive set of data for transitions between all levels
under consideration.
The primary goal of the present work, therefore, is to provide a coherent and consistent set of
data for electron collisions with Ca. From a fundamental point of view, it is also important
to estimate the likely uncertainties of the available theoretical data.
In addition to an accurate target representation, it is very important in this respect
to check the convergence of the close-coupling expansion,
particularly with respect to the contribution of the target continuum.
Based on the CCC results for the $4s4p\ ^1P^o$ excitation of Ca~\cite{Fursa2008} and our 
previous calculations for Be~\cite{Be2016} and Mg~\cite{Mg2017}, we expected a strong influence of the 
target continuum, which has not yet been explored for Ca to full extent.

The present calculations were performed with the BSR method (for an overview, see~\cite{BSR+}),
employing an extended version of the associated computer
code~\cite{BSR} that allows the inclusion of a sufficient number
of physical target states as well as continuum pseudo\-states
in the intermediate-energy regime.
Our previous calculations~\cite{Ca2006,Ca2007} contained only 39 physical target states.
We therefore decided to extend these calculations using modern computing facilities by additionally
including 444 continuum pseudo\-states. Using the same physical target states allows us to directly
estimate the effect of coupling to the target continuum on the calculated excitation cross sections.
Furthermore, the pseudo\-state approach also enables us to obtain ionization cross sections,
thereby making the scattering data consistent, effectively complete, and appropriate for plasma modelling.
Note that the ionization cross sections in neutral Ca are still not well known.
As discussed by Cvejanovic and Murray~\cite{Cvet2003}, there exists noticeable disagreement between
various experimental data, both in the energy dependence and the magnitude at the cross-section maximum.
Theoretically, single ionization of calcium by electron impact was so far only considered 
by perturbative methods.  Consequently, it is interesting to explore whether the non\-perturbative
convergent pseudo\-state approach can help in resolving the existing discrepancies.

\section{Computational methods}

The target-structure and collision calculations
in the present work were carried out in a similar manner
to our previous calculations of electron scattering from calcium at low incident energies~\cite{Ca2006,Ca2007}.
Consequently, we will only summarize the specific features for the present case below,
related to the pseudo\-state approach employed in the present calculations.

The target states of calcium were generated by combining the
multi-configuration Hartree-Fock (MCHF) and the \hbox{$B$-spline} box-based close-coupling (CC) methods~\cite{BSR_MCHF}.
In this approach, the target wave functions for neutral Ca, $3p^64snl$,  are expanded over the one-electron states 
of Ca$^{+}$, i.e., $3p^6nl$. Both valence and core-valence correlation are important for the
ground state and the low-lying excited states of Ca.
The core-valence correlations are included through the
core-excited configurations in the CI expansions of the ionic states, whereas the valence correlations
are accounted for through the mixing of different series $3p^6nln'l'$. Our final expansions include the $4snl$, $3dnl$,
$4pnl$, $5snl$, $4dnl$, and $5pnl$ series. The unknown radial functions $P_{n\ell}$ for the outer valence electron
were expanded in a \emph{B}-spline basis, and the corresponding
equations were solved subject to the condition that the wave
functions vanish at the boundary. The \emph{B}-spline coefficients for
the valence orbitals $P_{n\ell}$ were obtained by diagonalizing the
\emph{N}-electron atomic Hamiltonian. This scheme leads to  term-dependent valence orbitals,
which are optimized individually for the states of interest.
We also account for relaxation of the core orbitals caused by the deep
core penetration of the $3d$ orbital.

An alternative and widely-used method of incorporating core-valence correlation
is based upon applying a semi-empirical core-polarization potential
(as was done, for example, in the CCC calculations~\cite{Fursa2008}).
Although such a potential simplifies the calculations significantly and
can provide accurate excitation energies and oscillator strengths, the
question always remains how well the model potential can simulate {\it all\/}
core-valence correlation, including non\-local and non\-dipole contributions. In the present approach, we
therefore chose to include the core-valence correlation {\it ab initio\/} by adding
target configurations with an excited core.

The number of spectroscopic bound states that can be generated in the above scheme depends
on the size~$a$ of the $R$-matrix box.
We included 140 \emph{B}-splines of order~8 in the present calculations.
Choosing $a=80\,a_0$ (with $a_0 = 0.529\times 10^{-10}\,$m denoting the Bohr radius),
we obtained a good description for all low-lying states of Ca up to $4s8s\ ^1S$,
including some doubly-excited states of the $4p^2$ and $3d^2$.
As discussed in~\cite{Ca2006},
the above scheme provides a good target description regarding both the energy levels and the oscillator
strengths. The deviations in the level energies from the recommended values~\cite{NIST} are generally
less than 0.1~eV, except for the lowest $4s^2\ ^1S$ and
$4s4p\ ^3P^o$ states, where the correlation corrections are expected
to be most important. The accuracy of our binding
energies is close to that achieved by extensive MCHF calculations~\cite{Ca_MCHF},
and the current structure description represents a
substantial improvement over those used in previous
\emph{R}-matrix calculations. This is particularly noteworthy
for the $3dnl$ states: directly including the core relaxation for these states
drastically improves the corresponding binding energies.

The \hbox{$B$-spline} box-based close-coupling  method is also able to generate continuum pseudo\-states
that lie above the ionization threshold. The density and number of these states
again depend on the box radius and, to a lesser extent, on other \emph{B}-spline parameters,
such as their order and distribution on the grid.
The above approach is both a straightforward and general way to obtain the continuum
pseudo\-spectrum. It provides excellent flexibility by allowing us to vary the box radius
or to change the density of the \emph{B}-spline basis.
As will be illustrated below, including the continuum pseudo\-states is extremely important
to ensure the convergence of the final results for the excitation cross sections.

The scattering calculations were carried out by using a fully
parallelized version of the BSR complex~\cite{BSR}. To check the influence
of the target continuum, we set up two scattering models.
The first model, labeled BSR-39, includes 39 physical target states,
while the second model, labeled BSR-483, additionally contains 444 pseudo\-states
that cover the target continuum up to 20~eV above the first ionization threshold.  This model
includes all target states with orbital angular momenta $L \leq 4$, i.e., even $3p^6 4s5g\ ^{3,1}G$.
This scattering model also allows us to obtain the ionization cross sections.
The maximum number of scattering channels was 1,215. For a given \emph{B}-spline basis,
this number defines the size of the matrices involved, leading in the present case to
generalized eigenvalue problems with matrix dimensions up to 150,000.
Such large calculations require the use of super\-computers.

We calculated results for partial waves with total orbital angular
momenta up to $L_{max} = 30$ numerically. Overall, with the various
total spins and parities, this involved 124 partial waves.
The calculation for the external region was performed with a
parallelized version of the STGF program~\cite{stgf}.
We considered all transitions between physical states, with the principal difficulty being the
slow convergence of the partial-wave expansion for transitions
between close-lying levels. When necessary we employed a top-up procedure
based on the Coulomb-Bethe approximation~\cite{BS1987}.

\section{Results}

\subsection{Elastic cross sections}

\begin{figure}%[t!]
\includegraphics[width=0.45\textwidth]{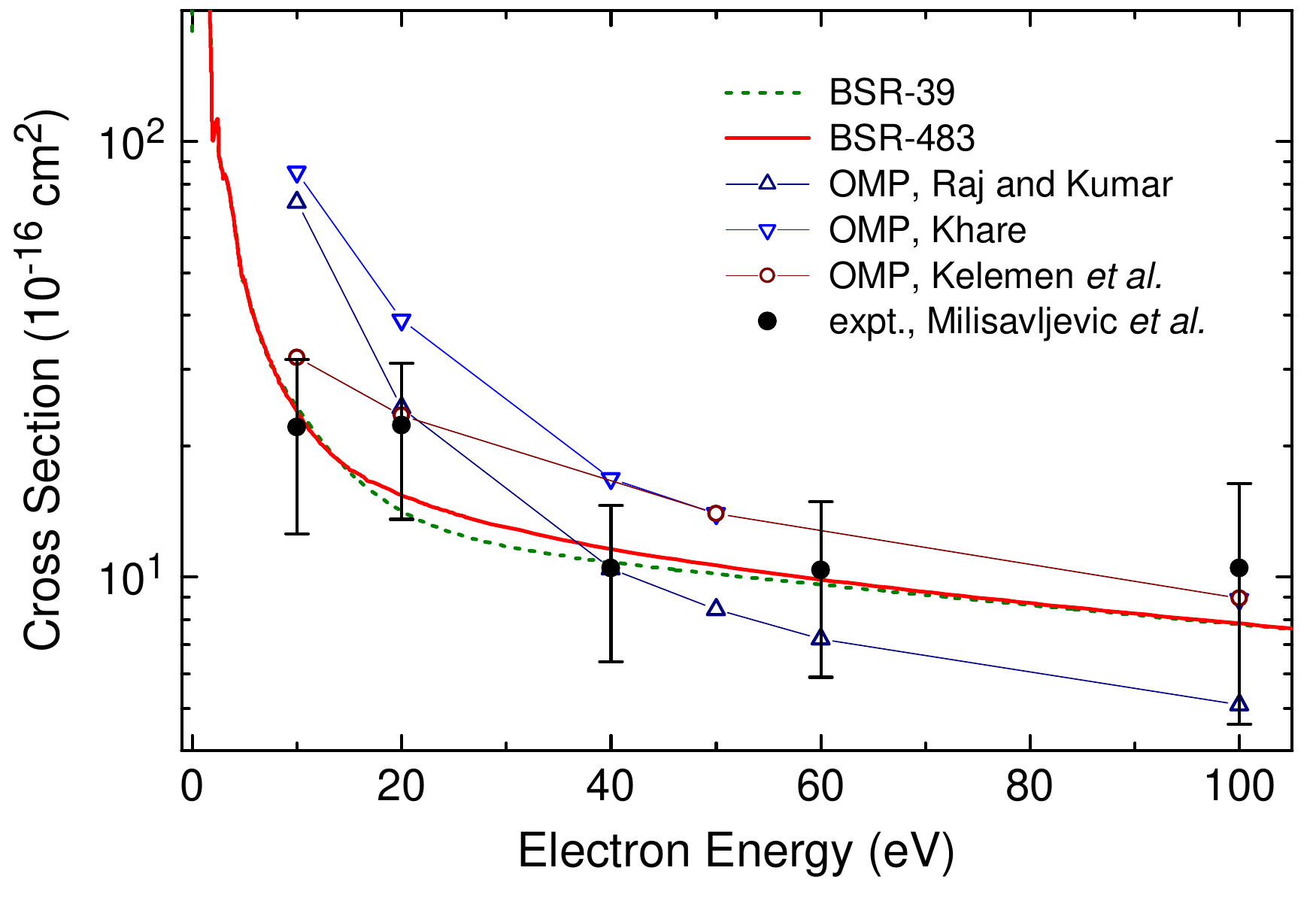}
\caption{\label{elastic} Cross section for elastic electron scattering
from calcium atoms at intermediate energies. The current BSR-39 and BSR-483 results are compared
with OMP calculations of Khare \emph{et al.}~\cite{Khare1985}, Kelemen \emph{et al.}~\cite{Kel1995},
Raj and Kumar~\cite{RK2007}, and with measurements of Milisavljevic \emph{et al.}~\cite{Mil2005}.}
\end{figure}

\begin{figure*}%[t!]
\includegraphics[width=0.8\textwidth]{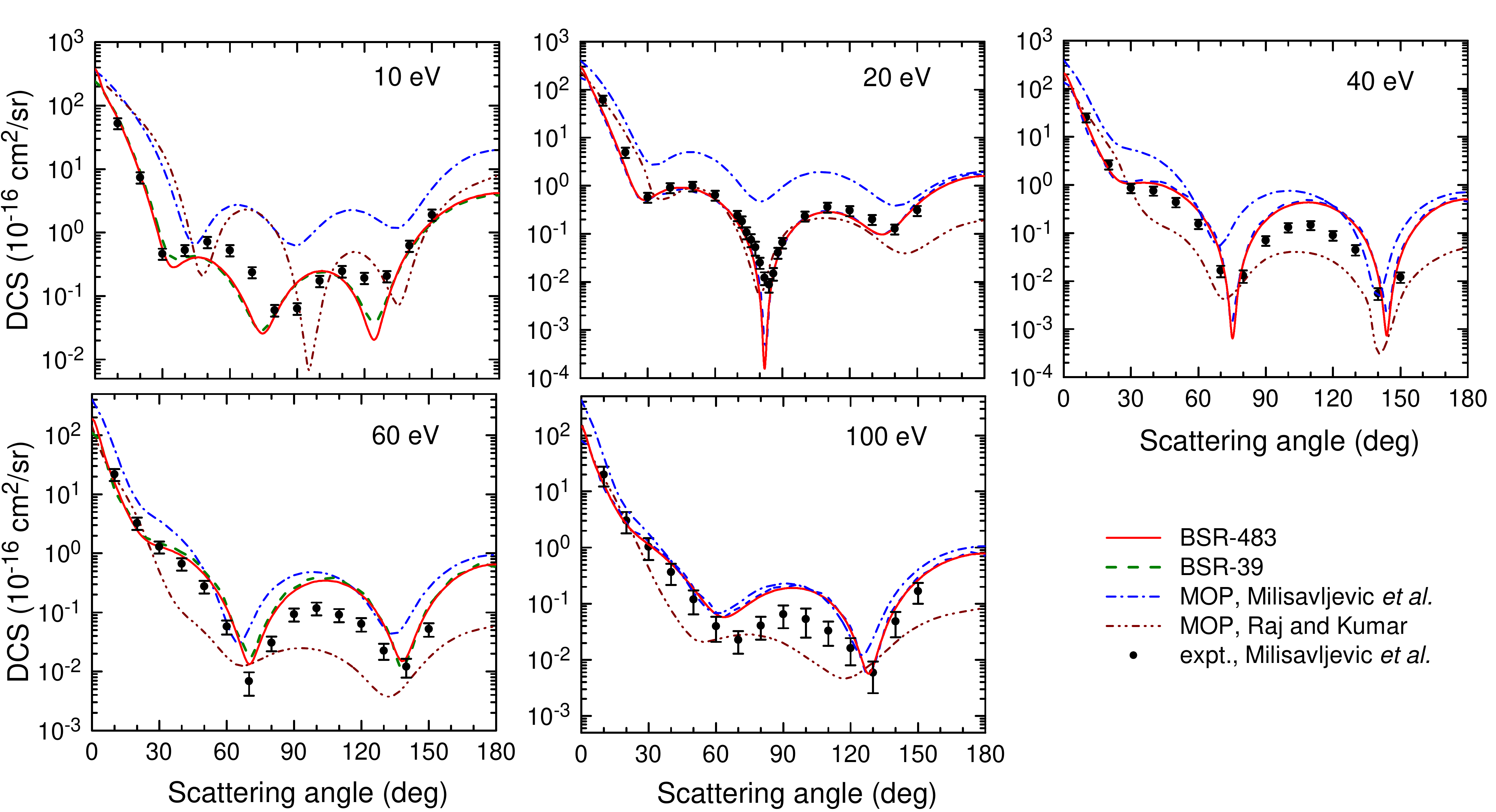}
\caption{\label{DCS_elastic} Angle-differential cross sections for elastic electron scattering from calcium atoms 
for incident electron energies of 10, 20, 40, 60 and 100~eV.
Our BSR results are compared with the experimental data of Milisavljevic \emph{et al.}~\cite{Mil2005}, as well as
with OMP calculations from the latter work and the OMP results of Rai and Kumar~\cite{RK2007}.}
\end{figure*}

We begin the discussion of our results with elastic scattering from calcium, which is
one of the most studied processes, both theoretically and experimentally.
The angle-integrated elastic cross section at intermediate energies is given in Fig.~\ref{elastic}.
We see close agreement between the BSR predictions and the experimental data of Milisavljevic \emph{et al.}~\cite{Mil2005}
for all energies within the given error bars,
except perhaps at except 20~eV,
where the experimental value deviates from the expected smooth energy dependence of the elastic cross section.
The small differences between the two sets of BSR results suggest that the effect of coupling to the target continuum for this process
is very small, essentially negligible.
Nevertheless, the good convergence in this case does not indicate that the calculations are trivial: for these  energies there are
many other calculations based on a variety of model potentials.  They yield rather different results
and agree with the measurements to a much lesser extent.
We hence conclude that the non\-perturbative R-matrix calculations are more reliable than previous predictions.

A comparison of angle-differential cross sections for elastic scattering is presented in Fig.~\ref{DCS_elastic}.
We obtain close agreement in both magnitude and angular dependence with the experimental DCS~\cite{Mil2005}.
The good agreement between the BSR-39 and BSR-483 results indicates, once again, 
that coupling to the target continuum does not change the angular dependence significantly.
The magnitude of the angle-integrated cross sections is almost completely determined by the small-angle region,
up to \hbox{20-25} degrees. The noticeable difference in the angle-integrated elastic cross sections
at 20~eV, as discussed above, is mainly caused by the difference in the DCS values at 10 and 20 degrees.
The absolute cross sections in~\cite{Mil2005} were obtained from the elastic-to-inelastic (to the
resonant $4^1P^o$ state) intensity ratio at $\theta=10$ degrees at each electron energy.
As a result, the accuracy of the measurements for the DCSs for excitation directly affect the accuracy of the elastic DCS.

A comprehensive list of references and a discussion of earlier calculations for the elastic DCS was
given by Milisavljevic \emph{et al.}~\cite{Mil2005}. In Fig.~\ref{DCS_elastic} we compare
only with a few somewhat recent calculations~\cite{Mil2005,RK2007}, which all used a model-potential
approach. This method provides a convenient and powerful tool for the
quantum-mechanical treatment of electron scattering by atoms and molecules in the
intermediate-energy region for complex atomic targets. It also offers a computationally simpler solution of the
differential equation (effectively for potential scattering) than direct close-coupling approaches 
such as the \emph{R}-matrix method used in the present work.  Because of these features, 
the model-potential method continues to attract considerable interest.
However, as seen from Figs.~\ref{elastic} and~\ref{DCS_elastic}, the accuracy of this approach is severely limited.

The angular dependence of the elastic DCS at 10~eV exhibits three minima, at 30$^\circ$, 70$^\circ$, and 120$^\circ$, respectively.
As the energy increases, these minima move and change:
the first minimum slowly transforms into a shoulder, the second moves towards
smaller scattering angles, and the third becomes more pronounced.
Generally, the calculated DCS curves show shapes similar to the experimental ones, except for the lowest energy of 10~eV where
the deviations are large.  This is not surprising, since channel coupling should be most important at this energy.
Also, the agreement with experiment regarding the absolute values is closest for the present \emph{R}-matrix calculations.
The optical potential used in~\cite{Mil2005} consists of the sum of static as well as local exchange and polarization
potentials. The polarization potential contains a semi-empirical cut-off parameter, which
was chosen to provide the best visual fit to the experimental DCS data at a
particular energy. The authors concluded that the best agreement was obtained just with the static approximation
while the static+exchange+polarization calculation generally gave larger DCS values.

Rai and Kumar~\cite{RK2007} also used a model-potential approach, and one purpose of their 
work was to assess the contribution of absorption effects. 
Their optical potential was represented by an energy-dependent
central, local and complex potential to simulate the static, exchange, polarization, absorption,
and spin-orbit interaction effects. This calculation, which takes all the above-mentioned effects into account,
represents the previously most comprehensive model potential 
for Ca atoms at intermediate energies.

As seen from Fig.~\ref{DCS_elastic}, the level of agreement between the absolute values of the cross
sections obtained by Rai and Kumar~\cite{RK2007}  and the experimental data varies depending
on the incident energy and the scattering angle. The authors concluded that the large error bars 
call for more accurate measurements to draw meaningful conclusions about the agreement of the absolute
DCS values obtained by theory and experiment. The good overall agreement of the measurement
with the present \emph{R}-matrix calculations, however, suggests that the measurement is sufficiently accurate 
to represent all main features in the DCS for elastic \hbox{e-Ca} scattering and, together with the present results,
can serve as a set of benchmark data to check the accuracy of other calculations.
Note that both OMP calculations (as well as many others) used a simple Hartree-Fock approximation for the static
potential. We suggest that this is insufficient in the case of Ca, where there is strong mixing between
the $4s^2$ and $4p^2$ configurations, and the 4s orbital itself is strongly affected by the $3p^6$ core polarization.

\subsection{Excitation cross sections}

\begin{figure}%[t!]
\includegraphics[width=0.45\textwidth]{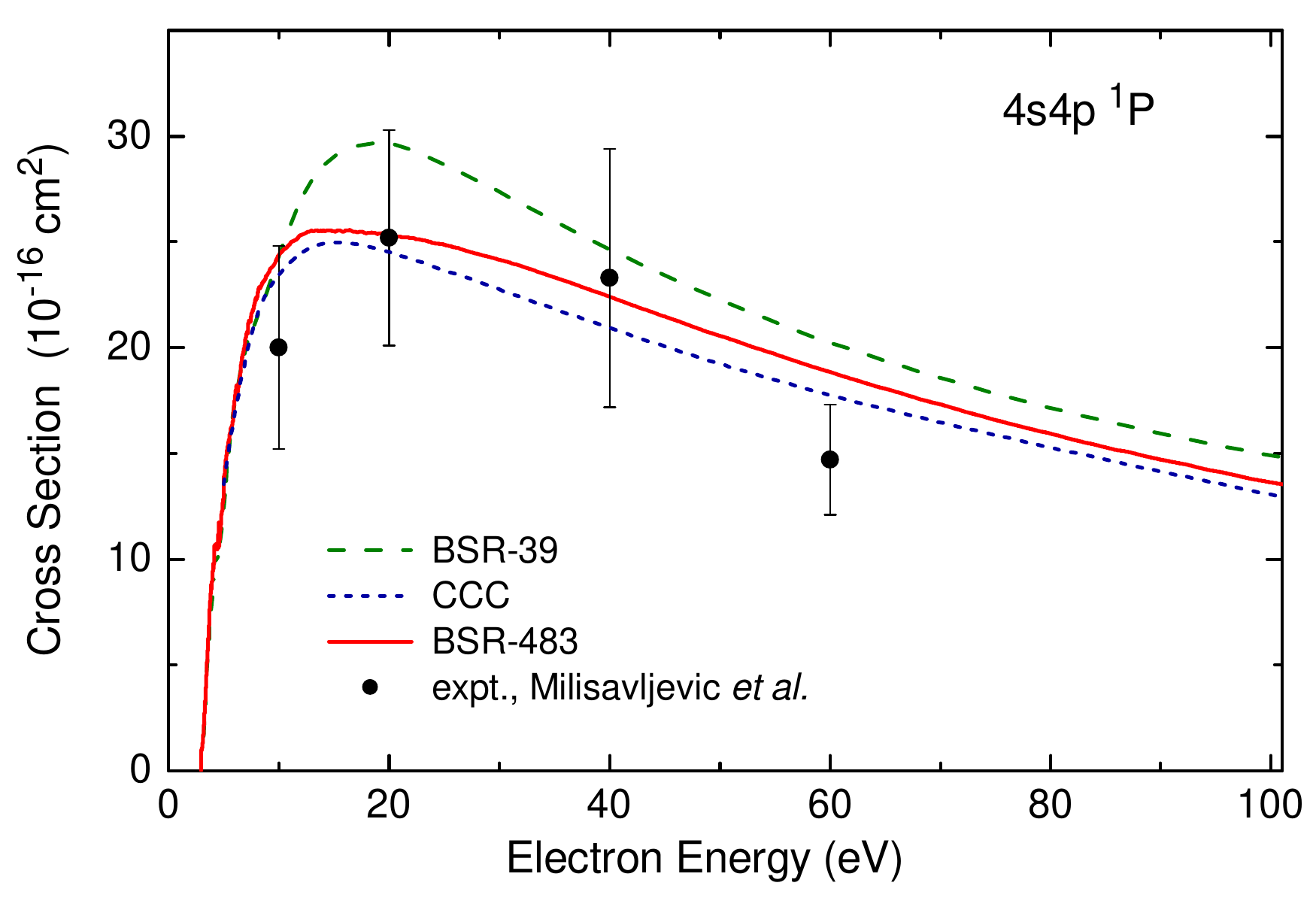}
\caption{\label{4s4p} Angle-integrated cross section for electron-impact excitation
of the $4s^2\ ^1S \rightarrow 4s4p\ ^1P^o$ resonance transition as a function of the incident electron energy.
Our BSR results are compared with the experimental data of Milisavljevic \emph{et al.}~\cite{Mil2004}
and CCC calculations of Fursa and Bray~\cite{Fursa2008}.}
\end{figure}

\begin{figure}%[t!]
\includegraphics[width=0.45\textwidth]{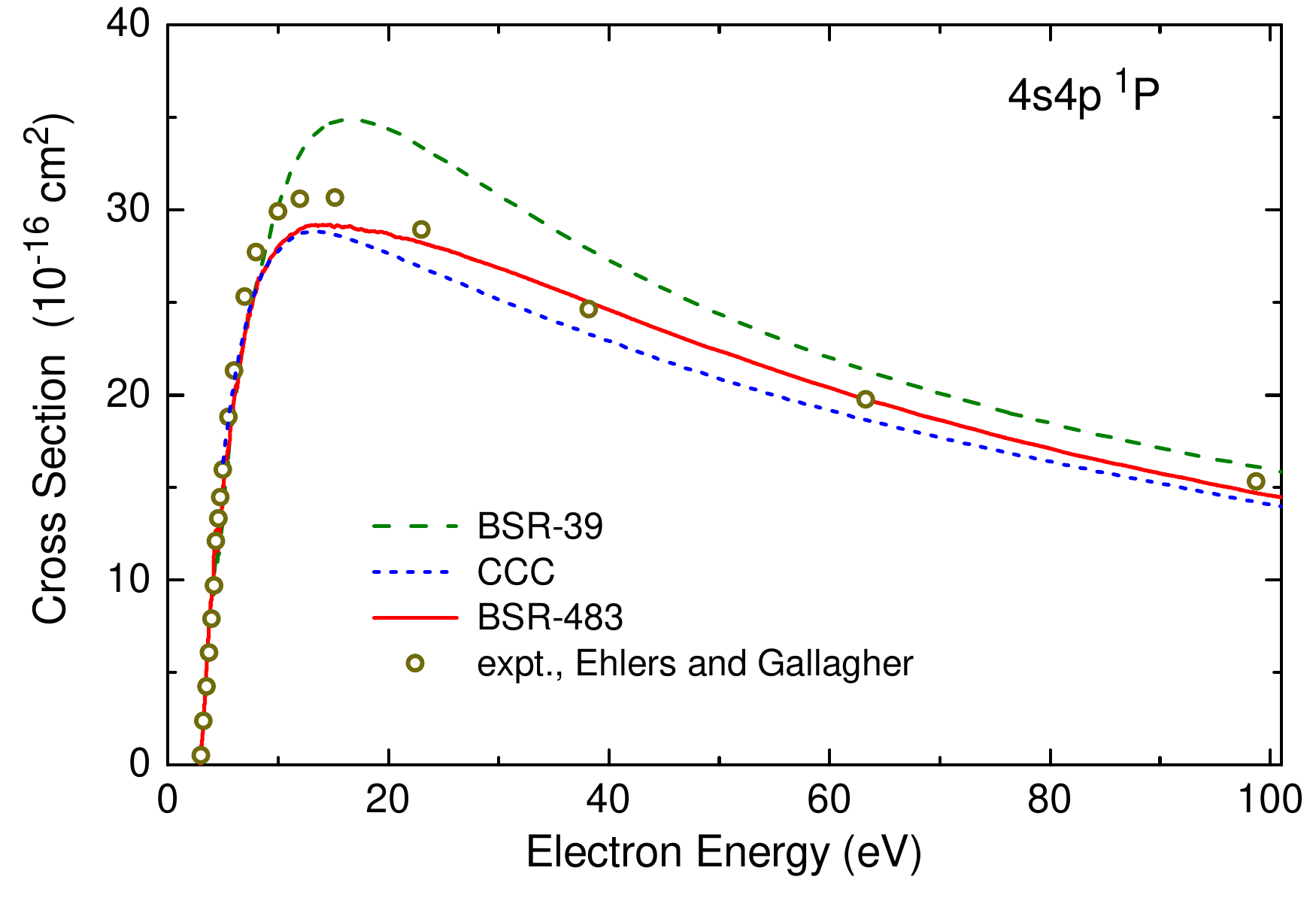}
\caption{\label{4s4p_apr} Apparent cross section for the $4s4p\ ^1P^o \rightarrow 4s^2\ ^1S$
spectral line as a function of the incident electron energy.
Our BSR results are compared with the experimental data of Ehlers and Gallagher~\cite{EG1973}
and CCC calculations of Fursa and Bray~\cite{Fursa2008}.}
\end{figure}

Figure~\ref{4s4p} shows the angle-integrated cross section for electron-impact excitation of the
$4s^2\ ^1S \rightarrow 4s4p\ ^1P^o$ resonance transition as a function of the incident electron energy.
The experimental points were obtained by integration of the angle-differential
measurements of Milisavljevic \emph{et al.}~\cite{Mil2004}.
From the available calculations we present only the most recent and most extensive calculations in the 
BSR-39 model~\cite{Ca2007}, the present pseudo\-state BSR-483 model, and the CCC 
results~\cite{Fursa2008}, which also include the effect of the target continuum.
Comparison of the two sets of BSR results with the same spectroscopic target wave functions allows us 
to assess the influence of the target continuum, whereas the comparison with the entirely 
independent CCC predictions provides an estimate for
the uncertainty in the theoretical data.
From this comparison we conclude that the target continuum has a noticeable influence 
even for this strong resonance transition, reducing the cross section at the maximum 
by $\sim$20\% and bringing the calculations into closer agreement with experiment. 
Some disagreement remains at 60~eV, but the experimental point seems low (see also Fig.~\ref{4s4p_apr} to be
discussed next).
The small differences between the BSR and CCC results is most likely due to the different target wave functions
used in these calculations.

Figure~\ref{4s4p_apr} compares the apparent 
cross section for the $4s4p\ ^1P^o \rightarrow 4s^2\ ^1S$
spectral line as a function of the incident electron energy.
Our BSR results are compared with the experimental data of Ehlers and Gallagher~\cite{EG1973}
and the CCC calculations of Fursa and Bray~\cite{Fursa2008}. The apparent cross sections include cascade
contributions from higher-lying states that may populate the radiating state. 
The contribution in this case is relatively small, 
not exceeding~15\%. Inclusion of the target continuum again improves the agreement with experiment, and 
we estimate the overall uncertainty
of the theoretical cross sections to be about~5\%. This illustrates the accuracy
that can be achieved by modern CCC or BSR codes for light quasi-two-electron atoms.

\begin{figure*}%[t!]
\includegraphics[width=0.8\textwidth]{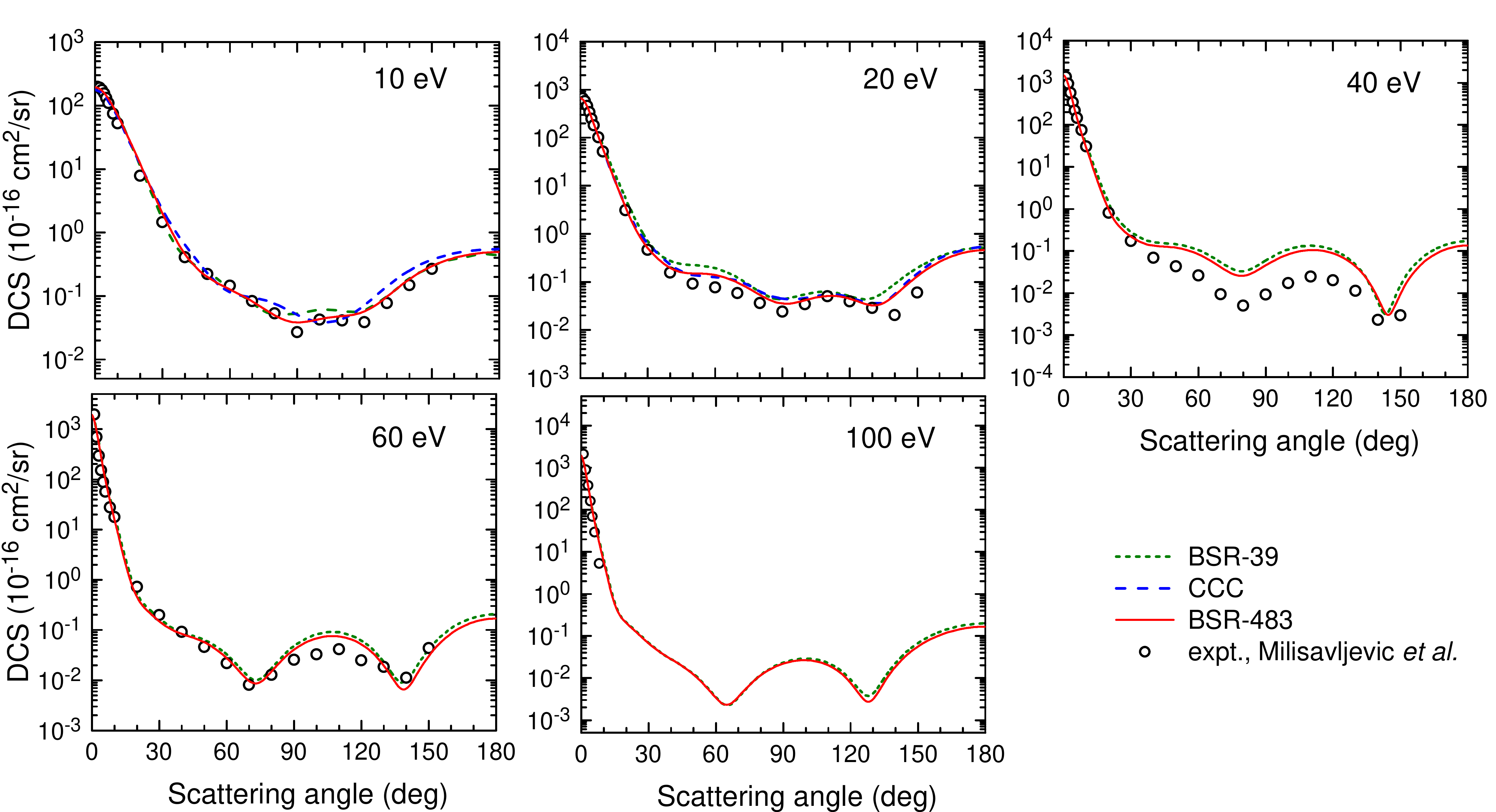}
\caption{\label{DCS_4s4p} Angle-differential cross sections for electron-impact excitation
of the $4s^2\ ^1S \rightarrow 4s4p\ ^1P^o$ resonance transition from calcium atoms
for incident electron energies of 10, 20, 40, 60 and 100~eV.
Our BSR results are compared with the experimental data of Milisavljevic \emph{et al.}~\cite{Mil2005}, as well as
with CCC calculations of Fursa and Bray~\cite{Fursa2008}.}
\end{figure*}

Angle-differential cross sections for electron-impact excitation
of the $4s^2\ ^1S \rightarrow 4s4p\ ^1P^o$ resonance transition
are presented in Fig.~\ref{DCS_4s4p}. We see very close agreement between
the BSR-39 and BSR-483 results for a wide range of scattering angles.
We conclude that the target continuum has a negligible influence on the angular dependence of the DCS
as a function of scattering angle.
This conclusion is similar to that drawn for the elastic DCS discussed above.
The difference in the angle-integrated cross sections comes mainly from the small-angle regime,
where the cross section rapidly changes by orders of magnitude.
Except at 40~eV, there is also close agreement with the experimental shape of the DCS curve.
Together with the close agreement between the CCC and BSR results at 10 and 20~eV, the comparison
confirms the high accuracy of the present results, which we consider converged for all current practical purposes.

\begin{figure*}%[t!]
\includegraphics[width=0.8\textwidth]{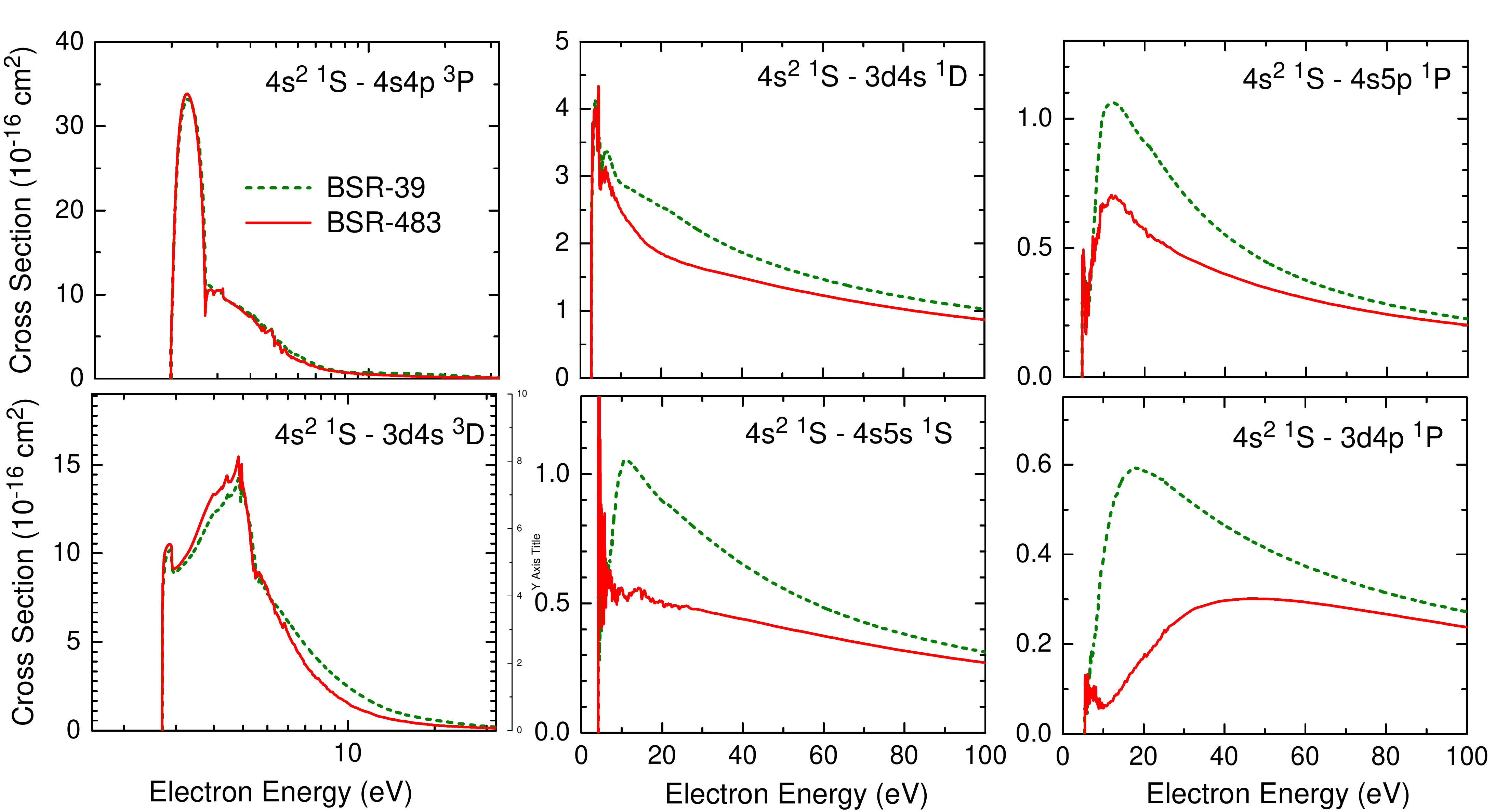}
\caption{\label{sct_nl1} Cross sections, as a function of collision energy, for
electron-impact excitation of the selected states of calcium from the $4s^2\ ^1S$
ground state.}
\end{figure*}

\begin{figure*}%[t!]
\includegraphics[width=0.8\textwidth]{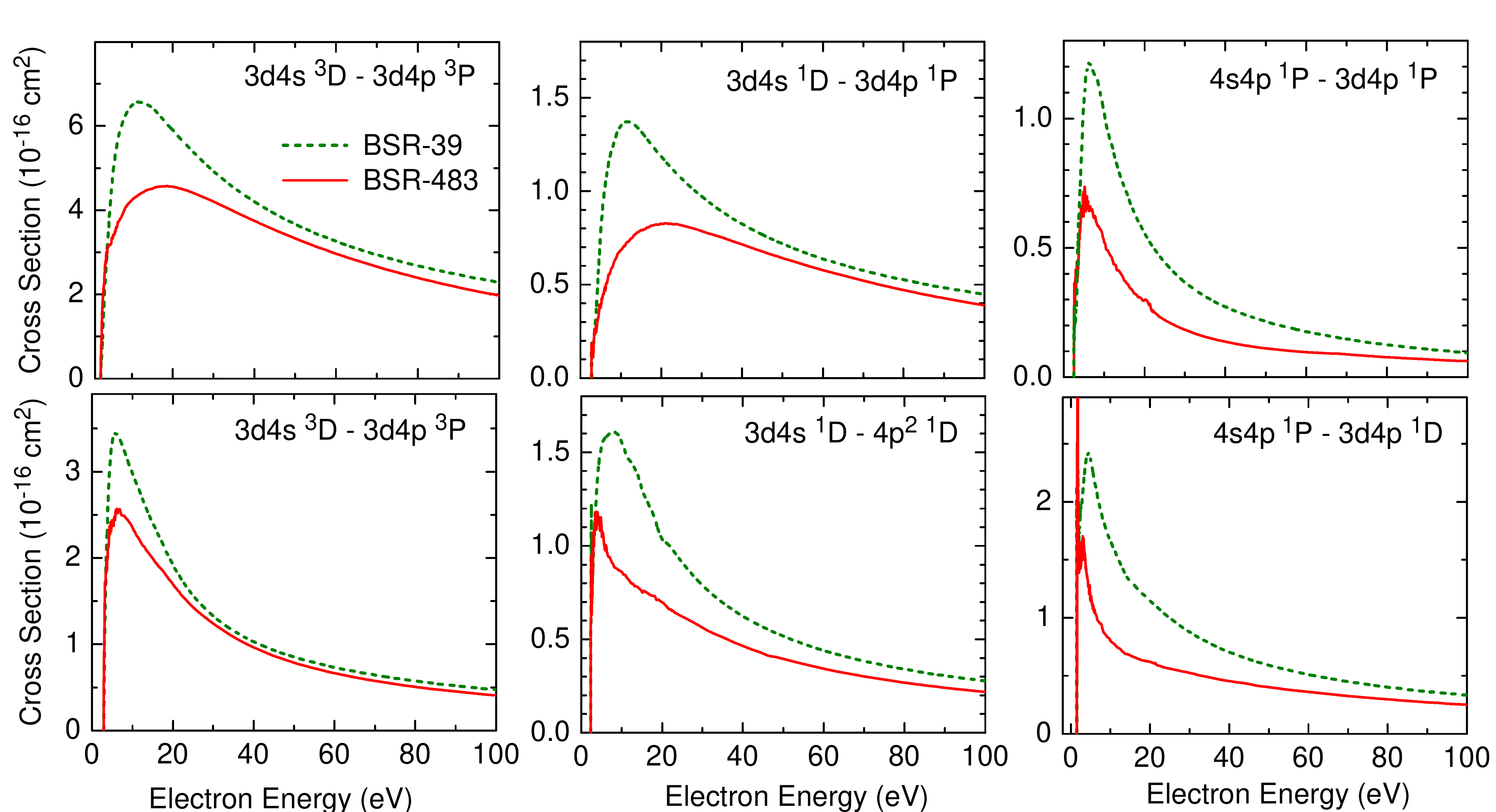}
\caption{\label{sct_nl2} Cross sections, as a function of collision energy, for
electron-induced transitions from selected excited states of atomic calcium.}
\end{figure*}

Figure~\ref{sct_nl1} exhibits a sample of results for excitation of different target states from the ground state.
The examples include different types of transitions, such as spin-forbidden exchange transitions
as well as monopole, dipole, and quadrupole transitions to states with the same total spin.
As a general trend, results for the exchange transitions converge quickly due to the short-range interaction, and  
they do not depend very much on the scattering model. Note that the exchange transitions to the $3d4s\ ^3D$ and $4s4p\ ^3P^o$ states
shown in the figure are intense, and near threshold their cross sections are comparable in magnitude with
those for the resonant dipole transition to the $4s4p\ ^1P^o$ state.
All spin-allowed transitions show a large influence of the target continuum, which decreases the cross sections
in the main near-threshold maximum. As expected, these corrections are most prominent
for weak two-electron transitions, such as $4s^2\ ^1S - 3d4p\ ^1P^o$, where the target continuum
corrections reach a factor of up to three and the peak energy is shifted significantly.

The same conclusions hold for transitions between excited states, which are shown in Fig.~\ref{sct_nl2}.
There are only very few other results at intermediate energies available for comparison.
We believe that all the predicted cross sections from our large-scale BSR-483 model 
(including those for transitions between excited states) are close to convergence,
but either experimental data or other independent calculations are needed for a reliable estimate of
potentially remaining uncertainties. For all calculated transitions, we expect the same level 
of accuracy as for elastic scattering and the resonance transition discussed above.

\subsection{Ionization cross sections}

\begin{figure}%[t!]
\includegraphics[width=0.45\textwidth]{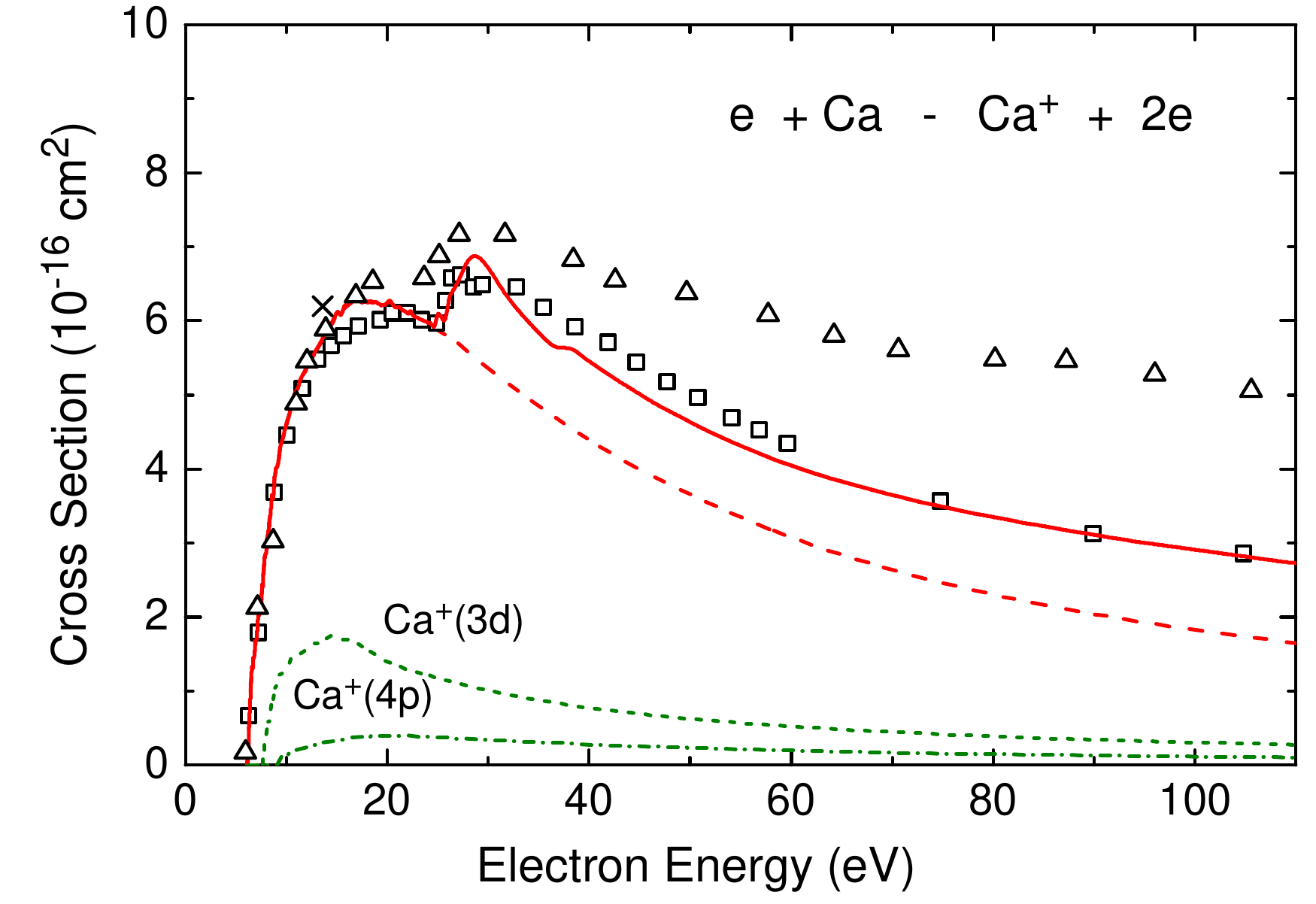}
\caption{\label{Ca_ion} Electron-impact ionization cross section for the
ground state of calcium. Measurements: squares, Okudaira~\cite{O1970}; crosses, Okuna~\cite{O1971},
triangles, Vainshtein \emph{et al.}~\cite{V1971}. Dashed line - BSR-483 results for the direct $4s$
ionization, solid line - plus $3p$-excitation. Also shown are results for ionization-excitation to the
$3p^63d$ and $3p^64p$ states of Ca$^+$.}
\end{figure}

The pseudo\-state approach also allows to generate ionization cross sections,
as the summation of all excitations to the continuum pseudo\-states.
Our ionization cross section is shown in Fig.~\ref{Ca_ion}.
The comparison with experiment is complicated here due to the fact that existing absolute measurements
were done for total ionization, where contributions of double ionization and inner-shell excitations are significant.
Our BSR-483 model only allows the consideration of direct $4s$-ionization.
For comparison with the measurements, we also added the excitation-autoionization contribution from the $3p$ subshell.
This contribution was obtained in separate calculations of the $3p$ excitation to the core-excited $3p^53d4s^2$ states,
assuming that the strong dipole $3p - 3d$ transitions provide the main contribution to this process.
As seen from Fig.~\ref{Ca_ion}, there is a good agreement with the relative measurements for single ionization
reported by Okudaira~\cite{O1970}, who used mass spectroscopy to select different ionization stages.
These data were normalized to our cross sections at 90~eV.
Comparison with the absolute measurements by Okuno~\cite{O1971} and Vainshtein et al.~\cite{V1971}
for total ionization of Ca suggests that our calculations may slightly underestimate the cross section at low impact energies.
We estimate the overall uncertainties of our cross sections for the single ionization of Ca to be about 10\%.

Our pseudo\-states approach, together with the projection technique to select different final ionic states~\cite{PRL},
is also able to consider the ionization + excitation process. As an example, Fig.~\ref{Ca_ion} also
presents the ionization-excitation cross sections to the $3p^63d$ and $3p^64p$ states of Ca$^+$.
These pathways yield relatively small contributions to the total ionization, and ionization-excitation to
other excited states of Ca$+$ was found to be negligible.

\subsection{Grand-total cross sections from ground states}

\begin{figure}%[t!]
\includegraphics[width=0.45\textwidth]{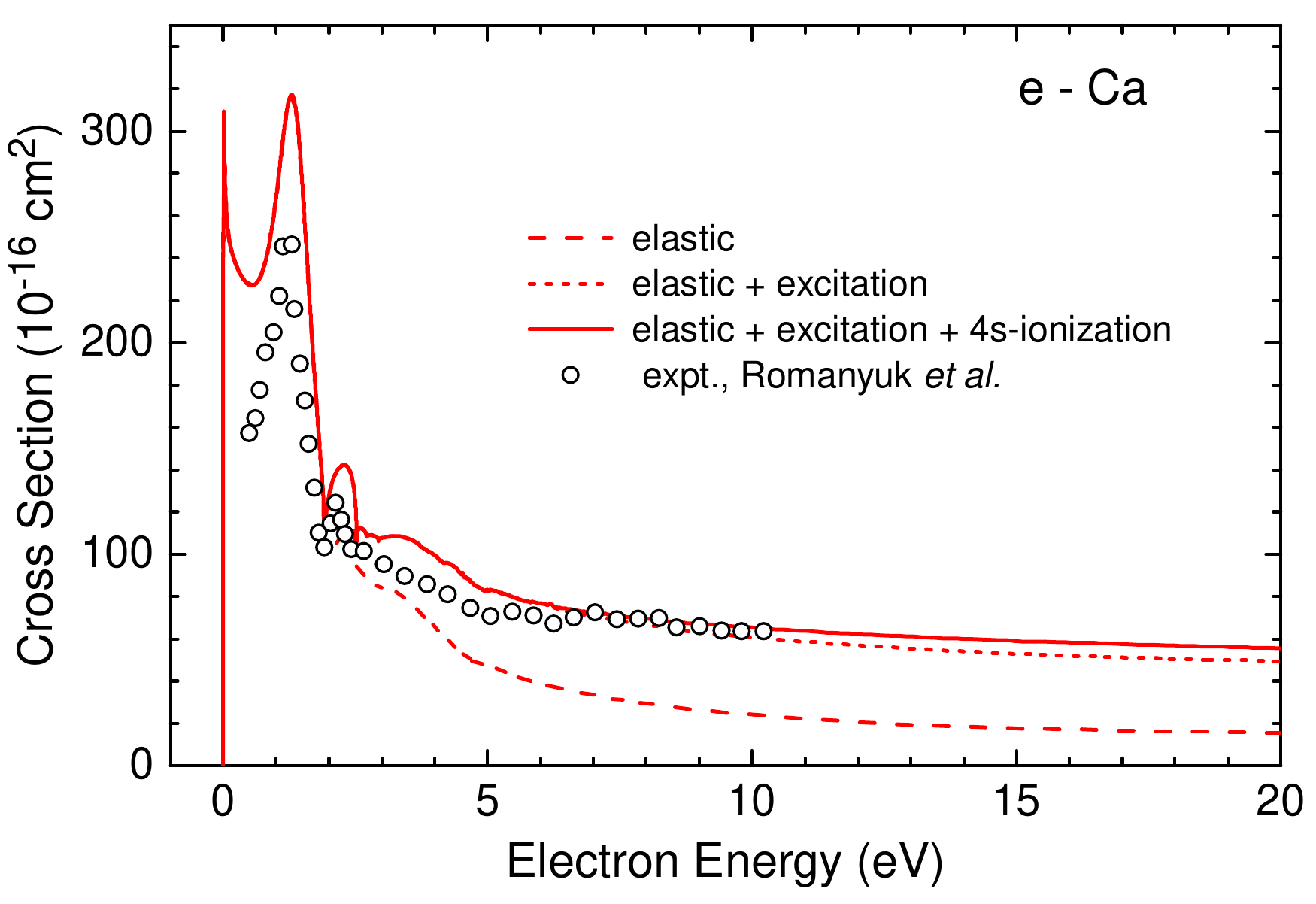}
\caption{\label{total} Elastic, elastic + excitation, and grand total cross section
for electron collisions with atomic calcium in the ground state, as obtained in the BSR-483 model.
Also shown are the experimental data of Romanyuk \emph{et al.}~\cite{Rom1992}.  }
\end{figure}

Our last comparison in Fig.~\ref{total} shows the grand total cross section
for electron collisions with calcium atoms in their $4s^2\ ^1S$ ground state.
This is the sum of the angle-integrated elastic, excitation, and
ionization cross sections. While the elastic cross section provides the
largest contribution at low energies, the contribution from excitation channels
becomes dominant already at energies above 5~eV.  The ionization
processes never contribute more than 10\% to the grand total
cross section. Note that the relative contribution of the elastic, excitation, and ionization processes
widely change from target to target.  In fact, ionization often dominates at higher energies.
We found such situations, for example, for electron scattering from atoms with an open $p$-shell.

There is a strong $d$-wave shape resonance at low energies, which was the subject our previous calculation~\cite{Ca2006}
in the BSR-39 model. The present calculations in the pseudo\-state approach, BSR-483, confirm
the previous results and conclusions. There is also good agreement in both the energy dependence and the magnitude 
of the grand total cross section
with the absolute measurements by Romanyuk \emph{et al.}~\cite{Rom1992}.
This further supports the accuracy of the present results.

\section{Summary}

We have carried out a detailed study of electron collisions
with neutral calcium, including elastic scattering, excitation,
and ionization processes from the ground and several excited
states. State-to-state excitation cross sections were obtained
for all transitions between the lowest 39 states of calcium.
We expect the cross sections presented here to be useful for
many practical applications.
While only a small number of selected results could be presented
in this paper, the entire dataset is available in electronic form
upon request.

The calculations were performed with the BSR code~\cite{BSR}.
The particular advantage of the approach is the possibility
to employ term-dependent non\-orthogonal one-electron
orbitals in the description of the target states.
This feature greatly improves the accuracy of the target description.
In particular, the present target wave functions
contain, fully {\it ab initio}, both valence and core-valence correlations, along
with the relaxation effects due to the significant penetration of
the $3d$ electron into the core.

The emphasis in the present calculations was placed on exploring
the influence of coupling to the target continuum.
The differences between the results from the \hbox{BSR-39} and \hbox{BSR-483} models
provide an indication regarding the convergence
of the close-coupling expansion for the problem at hand.
Overall, the influence of the target continuum was found to be significant
for most excitation cross sections, including even the strong resonance transition
to the $4s4p\ ^1P^o$ state, where coupling to the target continuum improved
the agreement with the available experimental data for the angle-integrated cross sections.
At the same time, the target continuum has a negligible influence
on elastic scattering and on the angular dependence (not the magnitude) of
the differential cross sections, for both excitation and elastic scattering.
We also showed that the present close-coupling calculations
yield much closer agreement with the available measurements than
all previous calculations using a model-potential approach.

Our pseudo\-state model, BSR-483, was also used to calculate
the electron-impact direct ionization cross section for the
calcium ground state. This fully non\-perturbative calculation
achieved good agreement with available experimental results
for the single ionization cross sections of Ca.
The excitation-autoionization contribution for ground-state ionization was
also found to be important. Finally, the grand total cross
section from the ground state, together with the contributions
from elastic scattering, excitation, and ionization, was
presented. We found large contributions from the excitation channels to the
total cross section at intermediate energies.

\begin{acknowledgments}
This work was supported by the United States National
Science Foundation under grant Nos.\ PHY-1403245, PHY-1520970, and PHY-1803844.
The numerical calculations were performed on STAMPEDE
at the Texas Advanced Computing Center. They were made
possible through the XSEDE allocation No.\ PHY-090031.
\end{acknowledgments}


\begin{thebibliography}{99}

\bibitem{Bark2016}  P. S. Barklem, Astron Astrophys Rev \textbf{24} (2016) 9.

\bibitem{Holw1972}  H. Holweger, Solar. Phys. \textbf{25} (1972) 14.

\bibitem{Thor2000}  P. Thoren, Astron. Astrophys. \textbf{358} (2000) L21.

\bibitem{Chmi2000}  Y. Chmielewski, Astron. Astrophys. \textbf{353} (2000) 666.

\bibitem{Chris2004} Christlieb, N., Gustafsson, B., Korn, A. J., et al.  ApJ, \textbf{603} (2004) 708.

\bibitem{Ca2006}    O. Zatsarinny, K. Bartschat, S. Gedeon, V. Gedeon, and V. Lazur,
                    Phys. Rev. A \textbf{74} (2006) 052708.

\bibitem{Mil2005}   S. Milisavljevic, D. Sevic, R. K. Chauhan, V. Pejcev, D. M. Filipovic,
                    R. Srivastava, and B. P. Marinkovic, J. Phys. B \textbf{38} (2005) 2371.

\bibitem{EG1973}    V. J. Ehlers and A.~Gallagher, Phys. Rev. A \textbf{7}, 1573 (1973);
                    {\it erratum:} Phys. Rev. A \textbf{9} (1974) 1026.

\bibitem{Mil2004}   S. Milisavljevic, D. Sevic, V. Pejcev, D.M. Filipovic, and B.P. Marinkovic,
                    J. Phys. B {\bf 37} (2004) 3571.

\bibitem{Chauhan2005} R. K. Chauhan, R. Srivastava, and A. D. Stauffer, J. Phys. B \textbf{38} (2005) 2385.

\bibitem{Kawazoe2005} S. Kawazoe T. Kai, R. K. Chauhan, R. Srivastava, and S. Nakazaki,
                      J. Phys. B \textbf{39} (2006) 493.

\bibitem{Ca2007}    O. Zatsarinny, K. Bartschat, L. Bandurina, and S. Gedeon, J. Phys. B \textbf{40} (2007) 4023.

\bibitem{Fursa2008} D. V. Fursa and I. Bray, J. Phys. B \textbf{41} (2008)  145206.

\bibitem{Ca1997}    I. I. Shafranyoshy, T. A. Snegurskaya, N. A. Margitich, S. P. Bogacheva,
                    V. I. Lengyel, and O. I. Zatsarinny,  J. Phys. B \textbf{30} (1997) 2261.

\bibitem{Muk2002}   K. Muktavat, R. Srivastava, and A. D. Stauffer, J. Phys. B \textbf{35} (2002 4797).

\bibitem{Sam2001}   A. M. Samson and K. A. Berrington, At. Data Nucl. Data Tables \textbf{77} (2001) 87.

\bibitem{Be2016}    O. Zatsarinny, K. Bartschat, D. V. Fursa, and I. Bray, J. Phys. B \textbf{49} (2016) 235701.

\bibitem{Mg2017}    P. S. Barklem, Y. Osorio, D. V. Fursa, I. Bray, O. Zatsarinny, K. Bartschat, and A. Jerkstrand,
                    A\&A \textbf{606} (2017) A11.

\bibitem{BSR+}      O. Zatsarinny and K. Bartschat, J. Phys. B \textbf{46} (2013) 112001.

\bibitem{BSR}       O. Zatsarinny, Comp.\ Phys.\ Commun.~\textbf{174} (2006) 273.

\bibitem{Cvet2003}  D. Cvejanovic and A. J. Murray, J. Phys. B \textbf{36} (2003) 3591.

\bibitem{BSR_MCHF}  O. Zatsarinny and C. Froese Fischer, Comp.\ Phys.\ Commun.~\textbf{180} (2009) 2041.

\bibitem{NIST}      {\tt http://physics.nist.gov/cgi-bin/AtData}.

\bibitem{Ca_MCHF}   C. Froese Fischer and T. Tachiev, Phys. Rev. A \textbf{68} (2003) 012507.

\bibitem{stgf}      N. R. Badnell, J. Phys. B \textbf{32}, 5583 (1999); see also
                    $http://amdpp.phys.strath.ac.uk/UK\_RmaX/codes.html$.

\bibitem{BS1987}    V. M. Burke and M. J. Seaton, J. Phys. B \textbf{19} (1986) L527.

\bibitem{Khare1985} S. P. Khare, A. Kumar, S. Vijay Shri, J. Phys. B \textbf{18} (1985) 1827.

\bibitem{Kel1995}   V. A. Kelemen, E. Yu. Remeta, and E. P. Sabad, J. Phys. B {\bf 28} (1995) 1527.

\bibitem{RK2007}    D. Raj and A. Kumar, J. Phys. B \textbf{40} (2007) 3101.

\bibitem{O1970}     S. Okudaira, J. Phys. Soc. Japan \textbf{29} (1970) 409.

\bibitem{O1971}     Y. Okuno, J. Phys. Soc. Japan \textbf{31} (1971) 1189.

\bibitem{V1971}     L. A. Vainshtein, V. I. Ochkur, V. I. Rakhovskii, and A. M. Stepanov,
                    Sov. Phys. JETP \textbf{34} (1972) 271.

\bibitem{PRL}       O. Zatsarinny and K. Bartschat, Phys. Rev. Lett. \textbf{107} (2011) 023203.

\bibitem{Rom1992}   N. I. Romanyuk, O. B. Shpenik, F. F. Papp, I. V. Chernysheva, I. A. Mandi,
                    V. A. Kelemen, E. P. Sabad, and E. Yu. Remeta, Ukr. Fiz. Zh. \textbf{37} (1992) 1639.

\end{thebibliography}
\end{document}